\documentclass[12pt]{article}
\usepackage{graphicx}
\usepackage{graphics}
\usepackage{fancybox}
\usepackage{amsmath,float,latexsym,psfrag,epsf,epsfig,amssymb,
chicago,rotating}
\usepackage{color}
\usepackage{url}

\usepackage{fullpage}
\usepackage[ruled,linesnumbered]{algorithm2e}
\usepackage{algorithmicx}

\newcommand{\comment}[1]{}

\newcommand{\bfY}{\hbox{\boldmath$Y$}}

\newcommand{\E}{{\mathbf{E}}}

\begin{document}

\begin{center}

{\Large \bfseries Evidence and Bayes factor estimation for Gibbs random fields}
\vspace{5 mm}

{\large N. FRIEL\footnote{\texttt{nial.friel@ucd.ie}}} \\
{\textit{School of Mathematical Sciences and Clique research cluster, CASL, \\ University College Dublin, Ireland.}}
\vspace{5 mm}

\today 

\vspace{5mm}

\end{center}

\bibliographystyle{mybib}

\begin{abstract}

\noindent Gibbs random fields play an important role in statistics. However they are complicated
to work with due to an intractability of the likelihood function and there has been much work devoted to finding
computational algorithms to allow Bayesian inference to be conducted for such so-called doubly intractable 
distributions. This paper extends this work and addresses the issue of estimating the evidence and Bayes factor 
for such models. The approach which we develop is shown to yield good performance.
Supplemental material for this article are available online. 


\end{abstract}

\section{Introduction}

Gibbs random fields find much use in statistics, for example, the widely used autologistic model \cite{bes74} in spatial
statistics, and the exponential random graph model in social network analysis \shortcite{rob:pat:kal:lus07}. The popularity 
of this class of statistical
model results from the fact that it is possible to build a joint model with complicated global dependencies based on local 
dependencies. Despite their wide use, Gibbs random fields present considerable difficulties from the point of 
view of parameter estimation, because the likelihood function is typically intractable for all but trivially small graphs. 
One of the earliest approaches to overcome this difficulty is the pseudolikelihood method \cite{bes72}, which approximates 
the joint likelihood function by the product of full-conditional distributions of all nodes, and so ignores dependencies 
beyond first order. Recently there has been much progress in the literature addressing this problem, based on the fact that 
it is often possible to simulate from the likelihood function, for example, the single auxiliary variable method 
\shortcite{mol:pet06}, the exchange algorithm \shortcite{Murray06} and approximate Bayesian computation model choice 
\shortcite{gre:mar09}, to name a few. 

Posterior parameter estimation for Gibbs random fields has been termed a doubly intractable problem because both the likelihood
function and also the posterior distribution are intractable. This paper tackles an additional layer of intractability by 
focusing on estimating the statistical evidence, that is, the integral of the un-normalised posterior distribution over the
parameters of the Gibbs random field and could be termed a triply intractable problem. 

This paper is organised as follows. Section~\ref{sec:gibbs_ran_fields} introduces Gibbs random fields and outlines the issue
of model choice for this class of statistical model. The exchange algorithm upon which the innovation in this paper is based
is explained in Section~\ref{sec:ex_alg}. Section~\ref{sec:popnMCMC} describes how the exchange algorithm can be extended to 
yield an estimate of the statistical evidence and the Bayes factor. Approximate Bayesian computation in the context of model 
choice for Gibbs
random fields is explained in Section~\ref{sec:ABC}. The performance of the Bayes factor estimation algorithm is illustrated in 
Section~\ref{sec:examples} using models from spatial statistics and statistical network analysis. Finally a summary and future
directions are presented in Section~\ref{sec:conclusions}.

\section{Discrete-valued Markov random fields}
\label{sec:gibbs_ran_fields}

Discrete Markov random fields play an important role in several areas of statistics including spatial statistics and social
network analysis. The autologistic model, popularised by Besag \citeyear{bes72} which has the Ising model as a special case, is
widely used in the analysis of binary spatial data defined on a lattice. The exponential random graph (or p$^*$) model is frequently
used to model relational network data. See \shortcite{rob:pat:kal:lus07} for an excellent introduction to this body of work. 

Despite their popularity, discrete valued Markov random fields (or Gibbs distributions) are very problematic to work with from
a inferential point of view. We will explain why below, and first begin with some preliminaries. Let $y=\{y_1,\dots,y_N\}$ denote 
realised data defined on a set of nodes $\{1,\dots,N\}$ of a graph, where each observed value $y_i$ takes values from some finite
state space. The likelihood of $y$ given a vector of parameters $\theta = (\theta_1,\dots,\theta_m)$ is defined as
\begin{equation}
 f(y|\theta) \propto \exp(\theta^T s(y)) := q(y|\theta),
\label{eqn:gibbs_like}
\end{equation}
where $s(y) = (s_1(y),\dots,s_m(y))$ is a vector of statistics which are sufficient for the likelihood. The constant of proportionality in 
(\ref{eqn:gibbs_like}), 
\[
 z(\theta) = \sum_{y\in Y} \exp(\theta^T s(y)),
\]
depends on the parameters $\theta$, and is a summation over all possible realisation of the Gibbs random field. Clearly, $z(\theta)$
is intractable for all but trivially small situations. 

One of the earliest approaches to overcome the intractability of (\ref{eqn:gibbs_like}) is the pseudolikelihood method \cite{bes72} 
which approximates the joint distribution of $y$ as the product of full-conditional distributions for each $y_i$,
\[
 f_{pseudo}(y) = \prod_{i=1}^n f(y_i|y_{-i},\theta),
\]
where $y_{-i}$ denotes $y\setminus\{y_i\}$. This approximation has been shown to lead to unreliable estimates of $\theta$, for example,
\cite{ryd:tit98}, \cite{fri:pet04}, \shortcite{fri:pet:rev09}. Monte Carlo approaches have also been exploited to estimate the intractable likelihood, for example the Monte
Carlo maximum likelihood estimator of  Geyer and Thompson \citeyear{gey:tho92}. More recently, auxiliary variable approaches have been
presented to tackle this problem through the single auxiliary variable method \shortcite{mol:pet06} and the exchange algorithm  
\shortcite{Murray06}. 

\subsection{Evidence estimation for MRFs is a triply intractable problem}

This paper explores the issue of estimation of the statistical evidence (or marginal likelihood) 
\[
 \pi(y) = \int_{\theta} f(y|\theta) \pi(\theta) d\theta. 
\]
The evidence can then be used to allow one to make probability statements about the model itself, which in practical terms amounts
to choosing which sufficient statistics to include in (\ref{eqn:gibbs_like}). Evaluating the statistical evidence is generally an
intractable calculation and one typically resorts to some form of Monte Carlo simulation in order to yield an estimate of it. 
For example, the tempered transition algorithm \cite{neal96}, annealed importance sampling \cite{neal01}, bridge sampling
\cite{men:won96} and power posteriors \cite{fri:pet08}, which is based on path sampling \cite{gel:men98}, all rely on the use
of a tempering or bridging scheme, typically transitioning from prior to posterior in order to estimate the evidence. For example, 
the power posterior method explores a tempered posterior distribution of the form
\begin{equation}
 \pi(\theta|y,t) \propto f(y|\theta)^t \pi(\theta),\;\;\; t\in{[0,1]},
\label{eqn:power_post}
\end{equation}
so that $\pi(\theta|y,t=0)$ and $\pi(\theta|y,t=1)$ correspond to the prior and posterior distribution, respectively. The 
evidence results from a discretised version of the identity 
\begin{equation}
\pi(y) = \int_0^1\E_{\theta|y,t}f(y|\theta)\;dt.
\label{eqn:power_post1}
\end{equation}

This article also uses tempering in a similar manner, however none of the approaches described above can be directly applied to 
posterior distributions with intractable likelihoods, since each method above (including equations (\ref{eqn:power_post}) and 
(\ref{eqn:power_post1})) assumes that 
the likelihood function $f(y|\theta)$, can be evaluated for any $\theta$. This is one of the reasons 
why estimating the evidence for doubly intractable distributions is a very challenging task, and is one which might be termed a 
triply intractable problem. This therefore motivates the development of an new strategy to overcome this intractability. The 
approach which we develop is based on the exchange algorithm which we now outline. 

\section{The exchange algorithm}
\label{sec:ex_alg}

Murray \textit{et al} \citeyear{Murray06} presented an algorithm to allow inference for doubly intractable distributions extending 
that presented in \shortcite{mol:pet06}. This algorithm samples from an augmented distribution
\begin{equation*}
  \pi(\theta',y',\theta|y) \propto
  f(y|\theta)\pi(\theta) h(\theta'|\theta) f(y'|\theta')
\end{equation*}		
whose marginal distribution for $\theta$ is the posterior of interest. Here $f(y'|\theta')$ is the same 
likelihood model on which $y$ is defined and $h(\theta'|\theta)$ is an arbitrary distribution for the augmented variable 
$\theta'$ which might depend on $\theta$, for example, a random walk distribution centred at $\theta$. The exchange
algorithm is described in detail below, where the output of the Markov chain at iteration $i$ is denoted by 
$(\theta^{(i)},\theta^{(i)'},y^{(i)'})$.

\begin{algorithm}[h]

Initialise $\theta^{(0)}, \theta^{(0)'}, y^{(0)'}$\;
\For{$i=1,\dots, I$}{
$\theta' \sim h(\cdot|\theta^{(i-1)})$\;
$y' \sim f(y|\theta')$\;
Propose to swap/exchange $\theta' \rightarrow \theta^{(i)}; \; \theta^{(i-1)} \rightarrow \theta^{(i)'}; y' \rightarrow y^{(i)'}$ and
with probability
\begin{equation}
 \alpha = \min\left( 1, \frac{f(y|\theta')}{f(y'|\theta')}
\frac{f(y'|\theta^{(i-1)})}{f(y|\theta^{(i-1)})}
\frac{\pi(\theta')\; h(\theta^{(i-1)}|\theta')}{\pi(\theta^{(i-1)})  h(\theta'|\theta^{(i-1)})} \right)
\label{eqn:exchange_accept}
\end{equation}
accept this move\;
Otherwise set $\theta^{(i-1)} \rightarrow \theta^{(i)}; \theta^{(i-1)'} \rightarrow \theta^{(i)'}; y^{(i-1)'} \rightarrow y^{(i)'}$.
}

\caption{Exchange algorithm\label{alg:exchange}}
\end{algorithm}

Taking a closer look at the acceptance ratio in the exchange algorithm (\ref{eqn:exchange_accept}), and assuming that $h$ is symmetric 
yields the expression
\begin{equation}
 \alpha = \min\left( 1, \frac{\pi(\theta')}{\pi(\theta^{(i)})}
 \frac{q(y|\theta')}{q(y|\theta^{(i)})} \frac{q(y'|\theta^{(i)})}{q(y'|\theta')} \right).
\label{eqn:exchange_ratio}
\end{equation}
If one were to apply a naive Metropolis-Hastings algorithm to sample from the target $f(y|\theta)\pi(\theta)$, by using a symmetric
proposal to move from $\theta$ to $\theta'$, this would yield the acceptance ratio
\begin{equation}
  \alpha = \min\left( 1, \frac{\pi(\theta')}{\pi(\theta)}
 \frac{ q_{\theta'}(y)}{q_{\theta}(y)} \frac{z(\theta)}{z(\theta')} \right)
\label{eqn:MH_ratio}
\end{equation}
which is intractable, due to the ratio $z(\theta)/z(\theta')$. Note that the intractable ratio $z(\theta)/z(\theta')$
in (\ref{eqn:exchange_ratio}) is replaced in (\ref{eqn:MH_ratio}) by the ratio $q_{\theta}(y')/q_{\theta'}(y')$ which
can be interpreted as importance sampling estimate of $z(\theta)/z(\theta')$ as pointed out in \shortcite{Murray06}.



A practical difficulty in implementing the exchange algorithm is the requirement to draw an exact sample 
$y'\sim f(\cdot|\theta')$. Perfect sampling is an obvious approach. A pragmatic alternative is to take a realisation 
from a long MCMC run with stationary distribution $f(y'|\theta')$ as an approximate draw and this is the approach
taken by \cite{cai:fri11}, \shortcite{cuc:mar09}, \cite{fri:pet11}, and a theoretical justification for this approximation is
given by \cite{ever12}.

\section{A population MCMC extension of the exchange algorithm}
\label{sec:popnMCMC}

Here we present a population-based MCMC extension of the exchange algorithm which will lead to realisations from
the posterior distribution $p(\theta|y)$. We will then illustrate how this algorithm can be modified to also
allow an unbiased estimate of the normalising constant of the likelihood, $\hat{z}(\theta^*)$, for each draw $\theta^*$
from the posterior distribution. Both of these will allow us to estimate the evidence $\pi(y)$.

A population MCMC approach often involves constructing an augmented target distribution
\[
 \pi_{t_0}(\theta_0|y) \times \dots \times \pi_{t_n}(\theta_n|y),
\]
where 
\[
 \pi_{t_i}(\theta_i|y) \propto f(y|\theta_i)^{t_i} \pi(\theta_i)
\]
and where $0=t_0<t_1<\dots <t_{n-1}<t_n=1$. The rationale for constructing such an augmented target distribution is that this facilitates
the states of the population $(\theta_0,\dots,\theta_n)$ to \textit{interact} with one another to slowly move from the prior, 
$\pi_{t_0}$ to the posterior, $\pi_{t_n}$. Here we extend this framework, by further augmenting the distribution of each chain in the
population. We define
\begin{equation}
 \pi_{t_0}(\theta_0,\theta'_0,y'_{01},\dots,y'_{0s}|y)\times \dots \times 
 \pi_{t_n}(\theta_n,\theta'_n,y'_{n1},\dots,y'_{ns}|y)
\label{eqn:popn_stationary}
\end{equation}
where
\[
 \pi_{t_j}(\theta_j,\theta'_j,y'_j|y) \propto 
   f(y|\theta_j)^{t_j}\pi(\theta_j)f(y'_j|\theta'_j)^{t_j}h(\theta'_j|\theta_0,\dots,\theta_j)
\]
and the marginal distribution for $\theta$ at temperature $t_n=1$ is the target distribution
of interest. 

The population exchange algorithm is described in detail in Algorithm~\ref{alg:popn_exchange}. 

\begin{algorithm}[h]

Initialise $(\theta_0^{(0)},\theta_0^{(0)'},y_{01}^{(0)'},\cdot,y_{0s}^{(0)'}),\dots,
(\theta_n^{(0)},\theta_n^{(0)'},y_{n1}^{(0)'},\cdot,y_{ns}^{(0)'})$\;
\For{$i=1,\dots, I$}{
\For{$j=0,\dots,n$}{
$\theta'_j \sim h(\cdot|\theta_0^{(i)},\dots,\theta_{j-1}^{(i)},\theta_j^{(i-1)})$\; 
Draw $y'_1,\dots,y'_s \sim f(y|t_j\theta'_j)$\;
Propose to swap/exchange $\theta'_j \rightarrow \theta_j^{(i)}; \; \theta_j^{(i-1)} \rightarrow \theta_j^{(i)'}; 
\; (y'_1,\dots,y'_s) \rightarrow (y_{j1}^{(i)'},\dots,y_{js}^{(i)'})$ 
and with probability
\[
 \alpha = \min\left( 1, \frac{f(y|\theta'_j)^{t_j}\; \pi(\theta'_j)\; f(y'_1|\theta_j^{(i-1)})^{t_j}\;
h(\theta_j^{(i-1)}|\theta_0^{(i)},\dots,\theta_{j-1}^{(i)},\theta'_j)}
{ f(y|\theta_j^{(i-1)})^{t_j}\; \pi(\theta_j^{(i-1)})\; f(y'_1|\theta'_j)^{t_j}\;
h(\theta'_j|\theta_0^{(i)},\dots,\theta_{j-1}^{(i)},\theta_j^{(i-1)}) } \right)
\]
accept this move \;
Otherwise set $\theta_j^{(i-1)} \rightarrow \theta_j^{(i)}; \; \theta_j^{(i-1)'} \rightarrow \theta_j^{(i)'}; 
\; (y_{j1}^{(i-1)'},\dots,y_{js}^{(i-1)'}) \rightarrow (y_{j1}^{(i)'},\cdot,y_{js}^{(i)'})$
}
\[
 \hat{z}(\theta_n^{(i)}) = \prod_{j=0}^{n-1}\left( \frac{1}{s}\sum_{k=1}^s 
   \frac{q(y_{jk}^{(i)'}|t_{j+1}\theta_{j+1}^{(i)})}{q(y_{jk}^{(i)'}|t_j\theta_j^{(i)})} \right)\times z(0)
\]

}

\caption{Population exchange algorithm\label{alg:popn_exchange}}
\end{algorithm}

A key stage in this algorithm is how the chains interact with one another, and of course this is much choice in
how the function $h(\theta'_j|\theta_0^{(i)},\dots,\theta_{j-1}^{(i)},\theta_j^{(i-1)})$ is designed. A simple approach is to 
center this proposal on the average of parameters from neighbouring chains, for example,
\[
 \theta'_j \sim N\left(\frac{\theta_{j-1}^{(i)}+\theta_j^{(i-1)}}{2},\sigma^2_{\theta}\right).
\]
Obviously more elaborate proposals, such as those based on the differential evolutionary Monte Carlo approach of
\cite{ter:vru08}, may also be employed. 

Note that a tempering scheme has also been proposed in Murray \textit{et al} \citeyear{Murray06}. There the purpose was to
show how the tempered transitions algorithm \cite{neal96} could be adopted to allow one to efficiently transition to a new
parameter by proposing a sequence of intermediate parameters which connect the current parameters to the proposed parameters.
In our set up, the rationale for augmenting the target distribution with a sequence of tempered distribution is also to
facilitate efficient mixing through the state space, but additionally to allow estimation of the normalisation constant of the
likelihood function, which we now explain. 

\subsection{Estimating the normalising constant $z(\theta)$}

The population exchange algorithm outlines how to generate draws from the stationary distribution in (\ref{eqn:popn_stationary}). 
Here we will illustrate how this algorithm can be modified to  yield an estimator of the normalising constant $z(\theta)$ for a 
draw $\theta$ from the target distribution at temperature $t_n=1$. 

At iteration $i$, when visiting chain $j$, line $5$ of Algorithm~\ref{alg:popn_exchange} yields a draw from the likelihood,
\[
 y'_j \sim f(y|\theta'_j)^{t_j} \propto q(y|t_j \theta'_j).
\]
Assume for the moment that the auxiliary parameter value $\theta_j'$ lead to a swap/exchange move that was accepted, so
that $\theta_j^{(i)} = \theta_j'$.
We view the algebraic part, $q(y|t_j \theta_j^{(i)})$ of $f(y|t_j \theta_j^{(i)})$ as an 
importance distribution for 'target' $f(y|t_{j+1}\theta_{j+1}^{(i)})$. The ratio of the un-normalised importance function and 
un-normalised target gives an unbiased estimate of the ratio of their corresponding normalising constants:
\[
 \frac{q(y_j|t_{j+1} \theta_{j+1}^{(i)})}{q(y_j|t_j \theta_j^{(i)})} \approx \frac{z(t_{j+1}\theta_{j+1}^{(i)})}{z(t_j\theta_j^{(i)})}.
\]
This can be improved by taking additional draws $y_{j1},\dots,y_{js} \sim f(y|t_j \theta_j^{(i)})$ at line $5$ of 
Algorithm~\ref{alg:popn_exchange}, giving an unbiased estimate of $z(t_{j+1}\theta_{j+1}^{(i)})/z(t_j\theta_j^{(i)})$,
\begin{equation}
\widehat{\left(\frac{z(t_{j+1}\theta_{j+1})}{z(t_j\theta_j)}\right)} =  
\frac{1}{s}\sum_{k=1}^s \frac{q(y_{jk}|t_{j+1} \theta_{j+1}^{(i)})}{q(y_{jk}|t_j \theta_j^{(i)})}.
\label{eqn:ratio:nc}
\end{equation}
Denoting the current state of the population at iteration $i$ to be
\[
 (\theta_0^{(i)},\theta_1^{(i)},\dots,\theta_n^{(i)}),
\]
the auxiliary draws can be used, via importance sampling to yield an estimate of
\begin{equation}
 \frac{z(\theta_n^{(i)})}{z(0)} = \frac{z(t_n\theta_n^{(i)})}{z(t_0\theta_0^{(i)})} =
   \frac{z(t_1\theta_1^{(i)})}{z(t_0\theta_0^{(i)})} \times\dots\times \frac{z(t_n\theta_n^{(i)})}{z(t_{n-1}\theta_{n-1}^{(i)})}.
\label{eqn:ratio:nc1}
\end{equation}
Here each ratio on the right hand side of (\ref{eqn:ratio:nc1}) is estimated by (\ref{eqn:ratio:nc}).
Note that in the case of an Ising model, $z(0)=2^N$, where, as before $N$ is the size of the lattice.

This implies that the parameter value, $\theta_j^{(i)}$, of chain $j$ at iteration $i$, has an 
associated set of draws $y_{j1},\dots,y_{js}$, for $i=1,\dots,s$, corresponding to additional auxiliary draws from 
$f(y|t_j\theta_j^{(i)})$. In terms of storage, it suffices to store the low dimension sufficient statistics, 
$s(y_{j1}),\dots,s(y_{js})$. 

\subsection{Estimating the evidence}

The developments in this paper have shown that
\begin{enumerate}
 \item the population exchange algorithm yields draws, $\{\theta^{(i)}\}$ from the posterior 
by transitioning from the prior;
 \item Additional auxiliary draws, at each iteration, can be used to give an estimate of $z(\theta^{(i)})$. 
\end{enumerate}
This output can be used to estimate the evidence in a very straightforward manner. 
Re-writing Bayes' theorem, it holds for any $\theta^*$ that,
\[
 \pi(y) = \frac{f(y|\theta^*)\pi(\theta^*)}{\pi(\theta^*|y)} = \frac{q(y|\theta^*)\pi(\theta^*)}{z(\theta^*)\pi(\theta^*|y)}.
\]
We can approximate this by
\begin{equation}
 \hat{\pi}_{\theta^*}(y) = \frac{q(y|\theta^*)\pi(\theta^*)}{\hat{z}(\theta^*)\hat{\pi}(\theta^*|y)}
\label{eqn:chib}
\end{equation}
where $\hat{\pi}(\theta^*|y)$ and $\hat{z}(\theta^*)$ are estimates of $\pi(\theta^*|y)$ and $z(\theta^*)$ 
using the output from the population exchange algorithm. Since in most applications $\theta$ is low dimensional (usually less than $5$),
in such situations it
should be feasible to use kernel density estimation to estimate $\pi(\theta^*|y)$. Furthermore, since we have estimates of the 
normalising constant
at every draw $\theta$ from $\pi(\theta|y)$, it makes sense to estimate $\hat{\pi}_{\theta}(y)$ in (\ref{eqn:chib}) for a 
range of high posterior $\theta$ draws and take an average of these estimates. For example, we consider the following estimate of the
evidence
\begin{equation}
 \hat{\pi}(y) = \frac{1}{r}\sum_{b=1}^r \hat{\pi}_{\theta_b}(y),
\label{eqn:chib_mean}
\end{equation}
where $\theta_1,\dots,\theta_r$ correspond to draws from a high posterior density region. 

\subsection{Extending the methodology to estimate Bayes factors}
\label{sec:extensions}

This methodology can be extended to more complications situations where there may be many parameters within 
each model, as is the case with the exponential random graph model which is used in social network analysis. It
is also possible to construct a MCMC framework which can be used to estimate Bayes factors directly, rather than
estimate the evidence for each pair of models separately. 
We outline below how the methods explored in this paper may be applied to estimate the Bayes factor between $2$ 
competing (nested) models
\[
 \pi(\theta_1|y,m_1)  \;\;\; \mbox{and} \;\;\;  \pi(\theta_1, \theta_2|y,m_2).
\]
Consider a distribution which is a geometric average of the un-normalised posterior distributions
for models $1$ and $2$:
\begin{equation}
 \pi_t(\theta_1,\theta_2|y) = \left\{ f(y|\theta_1)\pi(\theta_1) \right\}^{1-t} 
 \left\{ f(y|\theta_1, \theta_2)\pi(\theta_1, \theta_2) \right\}^t,
\label{eqn:BF_12}
\end{equation}
for $0\leq t\leq 1$. 
Suppose, for ease of exposition, that model $m_1$ has an associated parameter $\theta_1$ and corresponding sufficient
statistic $s_1(y)$, while model $m_2$ also involves $\theta_1$ and $s_1(y)$, and an additional parameter $\theta_2$ 
with sufficient statistics $s_2(y)$. In the context of discrete Markov random fields (\ref{eqn:BF_12}) can be expressed
as
\[
 \pi_t(\theta_1,\theta_2|y) = \frac{q(y|\theta_1,t \theta_2)}{z(\theta_1,t \theta_2)} \pi(\theta_1)^{1-t} \pi(\theta_1,\theta_2)^t,
\]
where $q(y|\theta_1,t\theta_2) = \exp\left\{\theta_1 s_1(y) + t\theta_2 s_2(y)\right\}$ and its normalising constant is written as 
$z(\theta_1,t \theta_2)$.
Similar to Section~\ref{sec:popnMCMC}, augmenting (\ref{eqn:BF_12}) with an auxiliary data $y'$ and parameter $\theta'$ as follows, 
will allow an MCMC kernel based on the exchange algorithm to apply,
\begin{eqnarray*}
 \pi_t(\theta_1,\theta_2,y',\theta'_1,\theta'_2|y) &=& \left\{ f(y|\theta_1)\pi(\theta_1)f(y'|\theta'_1) \right\}^{1-t} 
 \left\{ f(y|\theta_1, \theta_2)\pi(\theta_1, \theta_2)f(y'|\theta'_1,\theta'_2) \right\}^t h(\theta'_1,\theta'_2|\theta_1,\theta_2)\\
 &=& \frac{q(y|\theta_1,t \theta_2)}{z(\theta_1,t \theta_2)} \pi(\theta_1)^{1-t} \pi(\theta_1,\theta_2)^t 
  \frac{q(y'|\theta'_1,t \theta'_2)}{z(\theta'_1,t \theta'_2)} h(\theta'_1,\theta'_2|\theta_1,\theta_2).
\end{eqnarray*}
In an entirely similar manner to Section~\ref{sec:popnMCMC} a population MCMC strategy can be used, which in this situation 
will yield at the end of each full sweep:
\begin{enumerate}
 \item A draw $\theta_1^*$ from $\pi(\theta_1|y,m_1)$,
 \item a draw $(\theta_1^{\dagger}, \theta_2^{\dagger})$ from $\pi(\theta_1^{\dagger}, \theta_2^{\dagger}|y,m_2)$,
 \item an estimate of $z(\theta_1^{\dagger},\theta_2^{\dagger})/z(\theta_1^*)$.
\end{enumerate}
The output can be used to estimate:
\begin{eqnarray*}
 BF_{12} &=& \frac{ \pi(y|m_1) }{ \pi(y|m_2) } \\ 
 &=& \frac{ f(y|\theta_1^*,m_1)\pi(\theta_1^*|m_1) }{ \pi(\theta_1^*|y,m_1) } 
\frac{ \pi(\theta_1^{\dagger}, \theta_2^{\dagger}|y,m_2) } 
{ f(y|\theta_1^{\dagger}, \theta_2^{\dagger},m_2)\pi(\theta_1^{\dagger}, \theta_2^{\dagger}|m_2) }\\  
&=& \frac{ q(y|\theta_1^*,m_1)\pi(\theta_1^*|m_1) }{ z(\theta_1) \pi(\theta_1^*|y,m_1) } 
\frac{ z(\theta_1^{\dagger}, \theta_2^{\dagger}) \pi(\theta_1^{\dagger}, \theta_2^{\dagger}|y,m_2) } 
{ q(y|\theta_1^{\dagger}, \theta_2^{\dagger},m_2)\pi(\theta_1^{\dagger}, \theta_2^{\dagger}|m_2) },\\ 
\end{eqnarray*}
for any $\theta_1^*$ and $(\theta_1^{\dagger}, \theta_2^{\dagger})$.

\section{Approximate Bayesian Computation methods}
\label{sec:ABC}

The exchange algorithm and its population-based variant rely on repeated drawing realisations from the likelihood. This is
also the basis for approximate Bayesian computation (ABC), where the goal is also to allow one to make inference for posterior
distributions with intractable likelihoods. There has been considerable work in this area beginning with \shortcite{pri:sei99}, 
\shortcite{beau:zhang02}. An excellent review of the main developments in ABC is presented in \shortcite{mar:pud12}. It is 
important to note that the class of statistical models examined in this paper, Gibbs random fields, are defined in terms of
sufficient statistics, and therefore the issue of choice of summary statistics with which to compare pseudo-data and observed
data, and which most ABC algorithms are faced with, does not arise here. Didelot \textit{et al.} \citeyear{did:eve11} present an 
approach to estimate the evidence in the situation where one needs to choose appropriate summary statistics. In the context of 
model choice for Gibbs random fields, \shortcite{gre:mar09} present an ABC algorithm to allow one to performance inference across 
the joint model and parameter space. Specifically they make the observation that the vector $S(y) = \left(S_1(y),\dots,S_M(y)\right)$ 
of sufficient statistics for each of $M$ competing Gibbs random field models, with model prior $\pi(m)$, for $m=1,\dots,M$, is 
itself sufficient for the joint distribution over 
model indices and parameters. In turn this leads to the ABC model choice algorithm described in Algorithm~\ref{alg:abc}.

\vspace*{0.3cm}

\begin{algorithm}
Initialise $\theta^{(0)}$\;
\For{$i=1,\dots, I$}{
$m^* \sim \pi(m)$ (model prior) \;
$\theta^* \sim \pi(\cdot|m^*) $ (parameter prior)  \;
$y' \sim f(y|\theta')$\;
Compute the distance $\rho(S(y),S(y'))$\;
Accept $(\theta^*,m^*)$ if $\rho((S(y),S(y'))<\epsilon$
}
\caption{ABC Model Choice}
\label{alg:abc}
\end{algorithm}

\section{Examples}
\label{sec:examples}

\subsection{Ising model}

The Ising model, is defined on a regular lattice of size $m\times m'$, where $n=mm'$. It is used to model the spatial distribution 
of binary variables, taking values $-1,1$. It is defined in terms of a sufficient statistic,
\[
 s_1(y) = \sum_{j=1}^n \sum_{i\sim j} y_i y_j,
\]
where the notation $i\sim j$ means that the lattice point $i$ is a neighbour of lattice point $j$. 
Henceforth we assume that the lattice 
points have been indexed from top to bottom in each column and where columns are ordered
from left to right. For example, for a first order neighbourhood model where an interior point $y_i$ has neighbours 
$\{y_{i-m}, y_{i-1}, y_{i+1},y_{i+m}\}$. Along the edges of the lattice each point has either $2$ or $3$ neighbours. 
A second order model has in addition a further sufficient statistic, $S_2(y)$, which counts the interaction between  
additional neighbours corresponding to the $4$ nearest diagonally adjacent neighbours, with
corrections to the neighbourhood on the border of the lattice. Thus the two models under consideration, $m_1$ and $m_2$,
are expressed as
\begin{eqnarray*}
 \pi(\theta_1|y,m_1) &\propto& \frac{\exp\left\{ \theta_1 s_1(y) \right\} }{z(\theta_1)} \pi(\theta_1|m_1), \\
 \pi(\theta_1,\theta_2|y,m_2) &\propto& \frac{\exp\left\{ \theta_1 s_1(y)+\theta_2 s_2(y) \right\} }{z(\theta_1,\theta_2)} 
  \pi(\theta_1,\theta_2|m_2), \\
\end{eqnarray*}
so that, when $\theta_2=0$, $m_1$ is nested within $m_2$. 

%

Here we simulated $20$ realisations from a first-order Ising model and a further $20$ lattices realised from a second order Ising model, all 
defined on a $10\times 10$ lattice. The aim of the experiment is to estimate which model,
a first or a second order neighbourhood model, best explains each dataset, a posteriori. 
For this experiment the lattices are sufficiently small to allow a very accurate estimate of the evidence in the following manner. 
For model $m_2$, the normalising constant $z(\theta_1,\theta_2)$ can be calculated exactly for a grid of 
$\{\theta_1^{(i)}, \theta_2^{(i)}\}$ values, which can then be plugged 
into the right hand side of:
\[
 \pi(\theta_1^{(i)},\theta_2^{(i)}|y) \propto \frac{q(y|\theta_1^{(i)},\theta_2^{(i)})}{z(\theta_1^{(i)},\theta_2^{(i)})}
   \pi(\theta_1^{(i)},\theta_2^{(i)}),\;\; i=1,\dots,n. 
\]
Numerically integrating the right hand side over the fine grid yields an accurate estimate of $\pi(y|m_2)$. This serves as a ground truth against which 
to compare with the corresponding estimates of the model evidence. The evidence $\pi(y|m_1)$ may be similarly estimated. 
Exact calculation of $z(\theta_1,\theta_2)$ and the approach
described above to get a precise estimate of the evidence relies on algorithms developed in \cite{fri:rue07}. A Gaussian prior 
for $\theta_j$, for $j=1,2$, with zero mean and standard deviation $5$ was chosen. 

In terms of implementation, the Bayes factor Population exchange algorithm was run, transitioning from $m_1$ to $m_2$, for $20,000$ iterations and $5$ chains, 
where $200$ auxiliary iterations are used to generate an approximate draw from the likelihood, and $200$ additional draws used 
for the importance sampling estimates of ratio of the normalising constant at parameters between intermediate chains. The choice of the
temperature ladder $\{t_1,\dots,t_n\}$ is an important aspect of this algorithm. Here we define $t_i = (i/n)^5$, for $i=0,\dots,n$,
so that more temperatures are placed towards the prior than the posterior. 
Note that in common with other tempering methods such
as annealed important sampling \cite{neal01}, the choice of temperature placement is important. An approach towards an automatic
method for making this choice is presented in \shortcite{beh:fri:hur11} in the context of the tempered transitions algorithm \cite{neal96}.
Finally, the closest $100$ $\theta$ draws to the posterior mean of $\theta$ were used in (\ref{eqn:chib_mean}) to estimate $\pi(y)$.

The ABC model choice algorithm was run for $500,000$ draws from the joint model and parameter prior. At each iteration, $200$ 
auxiliary draws were used to approximate a draw from the likelihood. This resulted in approximately equivalent computing time to the
population exchange algorithm. In terms of choosing a tolerance level, we considered two
choices, $\epsilon$ corresponding to the $0.5\%$ quantile of the distances between pseudo-data and the observed data (corresponding
to $2,500$ draws), and also a more restrictive choice of $\epsilon$ corresponding to the $0.1\%$ quantile (resulting in $500$
draws). 

%

\begin{figure}
\begin{center}
\begin{tabular}{ccc}
\includegraphics[width=5.3cm]{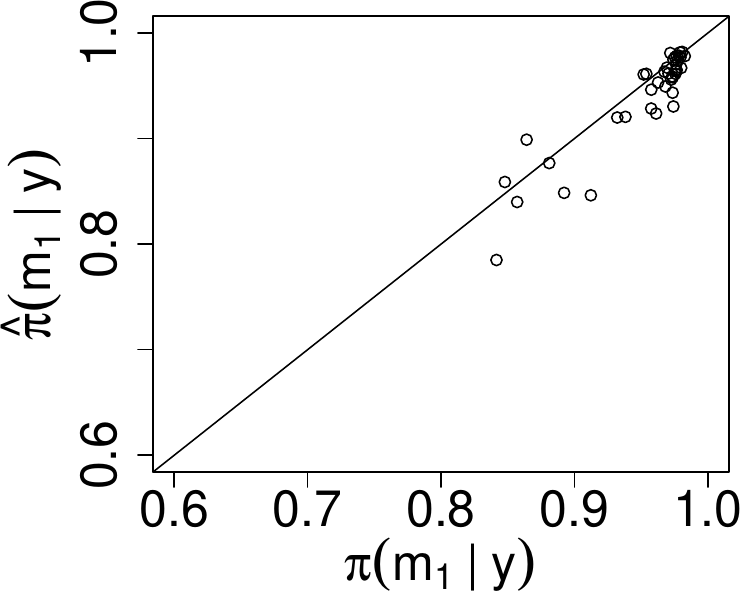} \hspace*{0.1cm} &
\includegraphics[width=5.3cm]{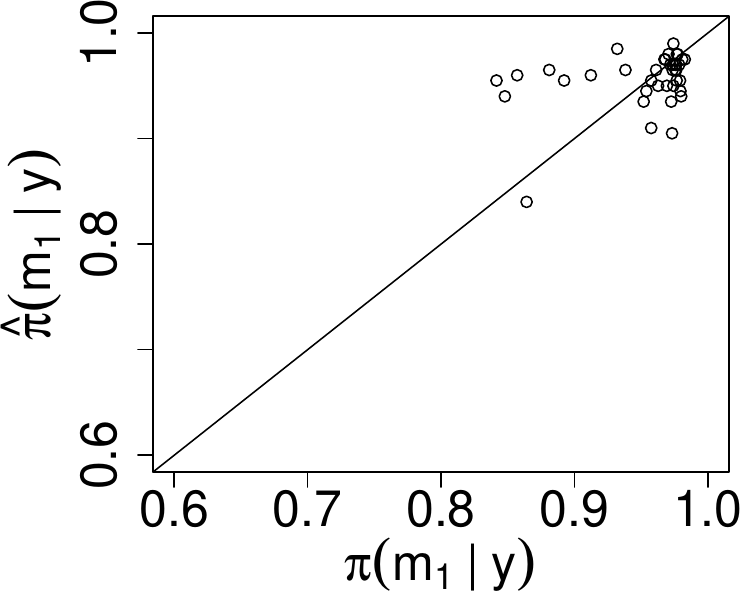} \hspace*{0.1cm} &
\includegraphics[width=5.3cm]{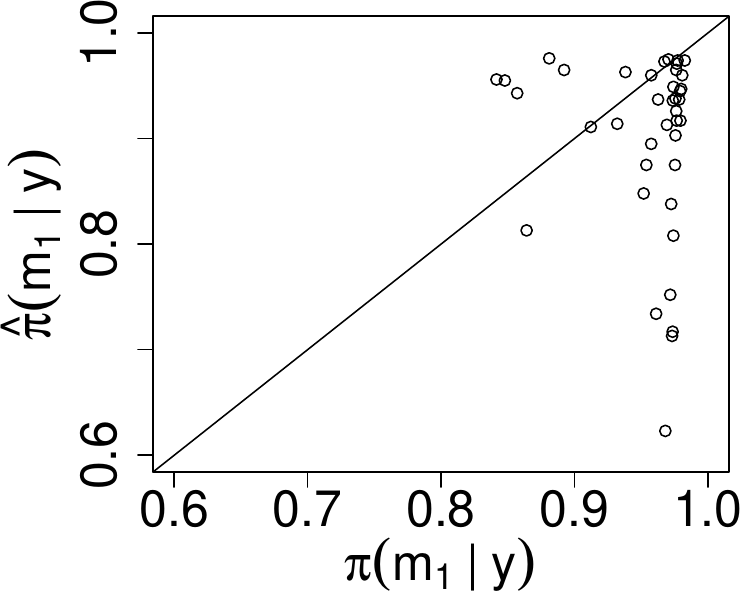} \\
(a) & (b) & (c) \\
\end{tabular}
\end{center}
\caption{(a) Evidence exchange (b) ABC (tolerance equal to the $0.1\%$ quantile) 
(c) ABC (tolerance equal to the $0.5\%$ quantile)}
\label{fig:p1}
\end{figure}

In order to compare the performance of the various algorithm, in Figure~\ref{fig:p1} we present a scatterplot of the true value of 
$\pi(m_1|y)$ against its estimated value $\hat{\pi}(m_1|y)$. 
We see that the evidence estimate based on the population exchange algorithm gives improved performance to the ABC model choice
algorithm with tolerance level $\epsilon$ corresponding to the $0.1\%$ quantile. However when the tolerance in the ABC algorithm is
set equal to the $0.5\%$ quantile, there is a much a large discrepancy between the true value of $\pi(m_1|y)$ and the estimated
$\hat{\pi}(m_1|y)$. 

\begin{figure}
 \begin{center}
 \includegraphics[width=9cm]{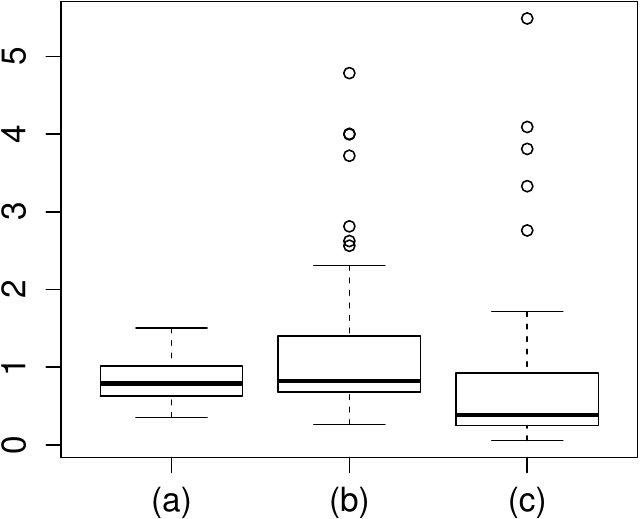}
\end{center}
\caption{Ratio of the true Bayes factor to the estimated Bayes factor for (a) Evidence exchange (b) ABC (tolerance set equal to the 
$0.1\%$ quantile) (c) ABC (tolerance set equal to the $0.5\%$ quantile)}
\label{fig:BF_plot}
\end{figure}

Finally, in Figure~\ref{fig:BF_plot} we display a boxplot of values of the ratio of the true Bayes factor to the estimated Bayes factor 
for each of the $40$ datasets. Again this suggests that the population exchange algorithm gives improved estimates of the Bayes factors 
compared to the ABC model choice algorithm with tolerance level equal to the $0.1\%$ quantile. While
the Bayes factor estimates based on using a tolerance level equal to a tolerance level of $0.5\%$ does not lead to efficient estimation of
the Bayes factor. 

%
%

\subsection{Exponential random graph model}

Here we explore how the methodology may be applied to the exponential random graph (ERG) model (or $p*$) \shortcite{rob:pat:kal:lus07} 
which is widely used in social network analysis. The ERG model is defined on a random adjacency matrix $\bfY$ of a graph on 
$n$ nodes (or actors) and a set of edges (dyadic relationships) $\{ Y_{ij}: i=1,\dots,n; j=1,\dots,n\}$ where $Y_{ij}=1$ if the 
pair $(i,j)$ is connected by an edge, and $Y_{ij}=0$ otherwise. An edge connecting a node to itself is not permitted so $Y_{ii}=0$. 
The dyadic variables maybe be undirected, whereby $Y_{ij}=Y_{ji}$ for each pair $(i,j)$, or directed, whereby a directed 
edge from node $i$ to node $j$ is not necessarily reciprocated. 

\begin{figure}
 \begin{center}
 \includegraphics[width=10cm]{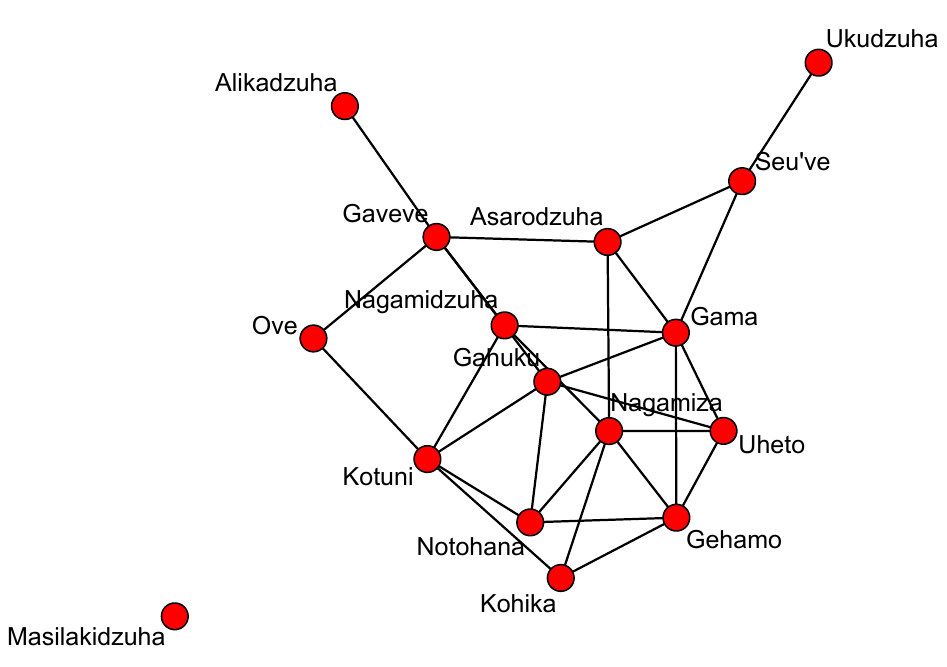}
\end{center}
\caption{Gamaneg graph}
\label{fig:gamaneg}
\end{figure}

The likelihood of an observed network $y$ is modelled in terms of a collection of sufficient statistics 
$\{s_1(y),\dots,s_k(y)\}$, each with corresponding parameter vector $\theta=\{\theta_1,\dots,\theta_k\}$,
\[
 f(y|\theta) \propto \exp\left\{ \sum_{l=1}^k \theta_l s_l(y) \right\}.
\]
For example, typical statistics include $s_1(y) = \sum_{i<j}y_{ij}$ and $s_2(y) = \sum_{i<j<k}y_{ik}y_{jk}$ which
are, respectively, the observed number of edges and two-stars, that is, the number of configurations of pairs of edges 
which share a common node. It is also possible to consider statistics which count the number of configuration of $k$  
edges which share a node in common, for $k>2$. 
A practical problem facing the practitioner is to understand which sufficient statistics to include in the model.
We view this is as a Bayesian model selection problem, and apply the developed methodology to the following network.
The Gamaneg network \cite{rea54}, displayed in Figure~\ref{fig:gamaneg} consists of $16$ sub-tribes of the Eastern central 
highlands of New Guinea. Here an edge represents an antagonistic relationship between two actors.

Here we consider two competing (nested) models. Model $m_1$ corresponds to the posterior distribution
\[
 \pi(\theta_1|y,m_1) \propto \exp\left\{ \theta_1 s_1(y) \right\} \pi(\theta_1),
\]
where $s_1(y)$ counts the total number of observed edge. Model $m_2$ corresponds to the posterior distribution 
\[
 \pi(\theta_1,\theta_2|y,m_2) \propto \exp(\theta_1 s_1(y) + \theta_2 s_2(y)) \pi(\theta_1,\theta_2),
\]
where, in addition to the $s_1(y)$ statistic, the two-star statistic $s_2(y)$, as defined above, is also included. 
For both models the prior distributions $\theta_1$ and $\theta_2$ are both set to be $N(0,5^2)$ distributions.

The population exchange algorithm outlined in Section~\ref{sec:extensions} was implemented with $10$ chains, for
$10,000$ overall iterations where $1,000$ auxiliary iterations were used to generate an approximate draw of a
network from the likelihood. A further $200$ draws were used for the importance sampling estimate of the ratio of
the normalising constant between successive temperatures. As before, a temperature schedule $t_i = (i/n)^5$, for 
$i=0,\dots,n$, was used. The algorithm was implemented in C and took approximately $20$ minutes on $3.33$Ghz processor 
with $4$Gb of memory. The algorithm yielded an estimate $BF_{12} = 37.499$, suggesting that there is considerably stronger 
support for model $m_1$ than for model $m_2$. 

\begin{figure}[h]
\begin{center}
\begin{tabular}{ccc}
  \includegraphics[width=5.5cm]{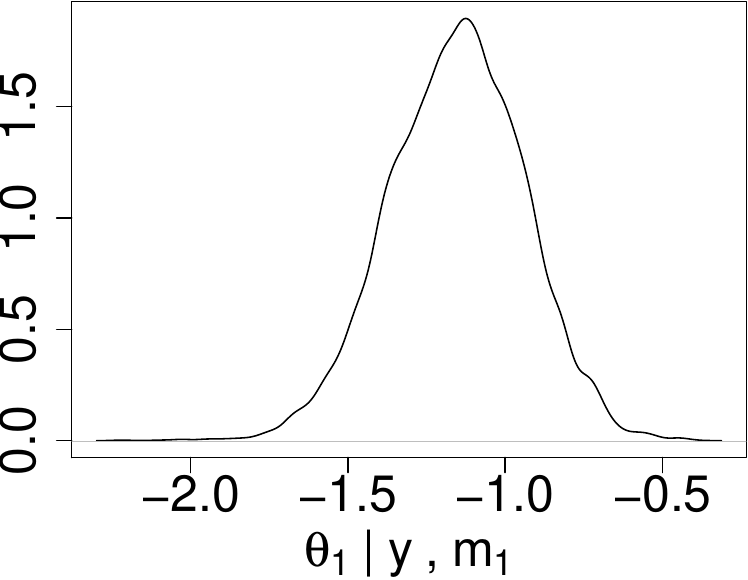} &
 \includegraphics[width=5.5cm]{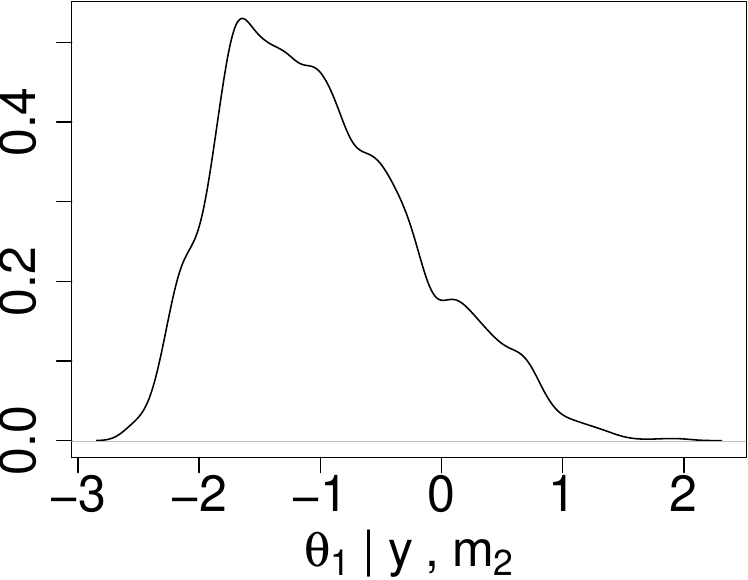} &
 \includegraphics[width=5.5cm]{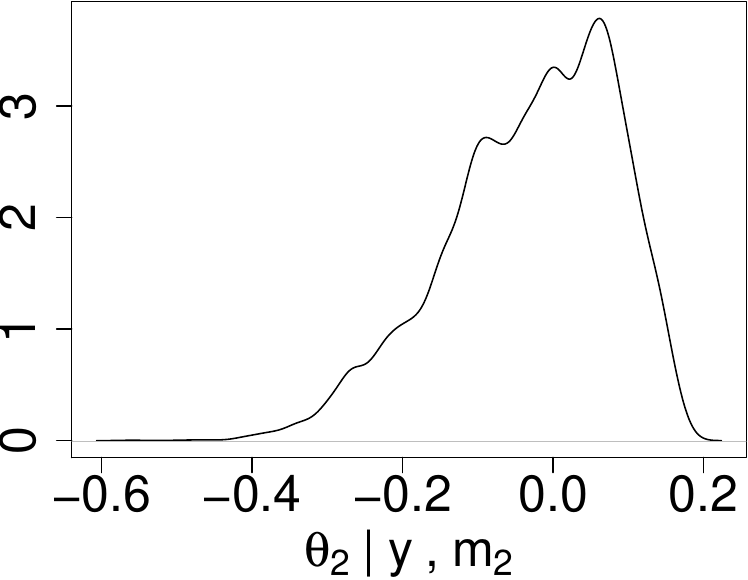} \\
(a) & (b) & (c)
\end{tabular}
\end{center}
\caption{Posterior density estimates for gamaneg network: (a) $\pi(\theta_1|y,m_1)$; (b) $\pi(\theta_1|y,m_2)$; 
(c) $\pi(\theta_2|y,m_2$).}
\label{fig:post_dens_ergm}
\end{figure}

An useful by-product of the population exchange algorithm is that draws at the first and last temperatures,
correspond to draws from model $m_1$ and $m_2$, respectively. The estimated posterior densities of parameters
from each model is displayed in Figure~\ref{fig:post_dens_ergm}. 

The posterior density for parameter $\theta_2$ in model $m_2$, in Figure~\ref{fig:post_dens_ergm}, illustrates that 
values of $\theta_2$ close to $0$ have strong support, a posteriori, which is consistent with most posterior support given to model $m_1$. 

\section{Summary}
\label{sec:conclusions}

Calculating the statistical evidence for Gibbs random fields is a challenging task. This article has proposed a novel
approach to this end. The experiments carried out here suggest that this method has some promise, and the next stage
will be to explore how this methodology extends to more complicated situations, such as hidden Markov random fields
and to larger exponential random graph models, both in the size of the graph, and the number of parameters in the 
model. 

\paragraph*{Acknowledgements:} Nial Friel's research was supported by a Science Foundation Ireland 
Research Frontiers Program grant, 09/RFP/MTH2199.

\section*{Supplemental Materials}

\begin{description}
 \item[C code:] The supplemental files for this article include C programs which can be used to
replicate the Ising model study and exponential random graph example in Section~\ref{sec:examples} of this article. 
Please see the file \texttt{README.txt} contained within the accompanying tar file for more details. 
\end{description}

\bibliography{hmrf}

\end{document}